\documentstyle[12pt]{article}

\def\ie{\hbox{\it i.e.}{}}

\catcode`\@=11

\@addtoreset{equation}{section}
\def\theequation{\thesection\arabic{equation}}

\def\@normalsize{\@setsize\normalsize{15pt}\xiipt\@xiipt
\abovedisplayskip 14pt plus3pt minus3pt%
\belowdisplayskip \abovedisplayskip
\abovedisplayshortskip  \z@ plus3pt%
\belowdisplayshortskip  7pt plus3.5pt minus0pt}
\def\small{\@setsize\small{13.6pt}\xipt\@xipt
\abovedisplayskip 13pt plus3pt minus3pt%
\belowdisplayskip \abovedisplayskip
\abovedisplayshortskip  \z@ plus3pt%
\belowdisplayshortskip  7pt plus3.5pt minus0pt
\def\@listi{\parsep 4.5pt plus 2pt minus 1pt
            \itemsep \parsep
            \topsep 9pt plus 3pt minus 3pt}}

\def\underline#1{\relax\ifmmode\@@underline#1\else
        $\@@underline{\hbox{#1}}$\relax\fi}
\@twosidetrue
\relax

\catcode`@=12

\catcode`\@=11

\def\section{\@startsection{section}{1}{\z@}{3.5ex plus 1ex minus
   .2ex}{2.3ex plus .2ex}{\large\bf}}
\def\thesection{\arabic{section}.}

\def\ps@headings{\def\@oddfoot{}\def\@evenfoot{}
\def\@oddhead{\hbox{}\hfill
        \makebox[.5\textwidth]{\raggedright\ignorespaces --\thepage{}--
        \hfill }}
\def\@evenhead{\@oddhead}
\def\subsectionmark##1{\markboth{##1}{}}
}

\ps@headings

\catcode`\@=12

\relax
\def\r#1{\ignorespaces $^{#1}$}
\def\figcap{\section*{Figure Captions\markboth
        {FIGURECAPTIONS}{FIGURECAPTIONS}}\list
        {Fig. \arabic{enumi}:\hfill}{\settowidth\labelwidth{Fig. 999:}
        \leftmargin\labelwidth
        \advance\leftmargin\labelsep\usecounter{enumi}}}
 \relax
\def\tablecap{\section*{Table Captions\markboth
        {TABLECAPTIONS}{TABLECAPTIONS}}\list
        {Table \arabic{enumi}:\hfill}{\settowidth\labelwidth{Table 999:}
        \leftmargin\labelwidth
        \advance\leftmargin\labelsep\usecounter{enumi}}}
 \relax
\def\reflist{\section*{References\markboth
        {REFLIST}{REFLIST}}\list
        {[\arabic{enumi}]\hfill}{\settowidth\labelwidth{[999]}
        \leftmargin\labelwidth
        \advance\leftmargin\labelsep\usecounter{enumi}}}
 \relax

\catcode`\@=11

\def\marginnote#1{}
\newcount\hour
\newcount\minute
\newtoks\amorpm
\hour=\time\divide\hour by60
\minute=\time{\multiply\hour by60 \global\advance\minute by-
\hour}
\edef\standardtime{{\ifnum\hour<12 \global\amorpm={am}%
    \else\global\amorpm={pm}\advance\hour by-12 \fi
    \ifnum\hour=0 \hour=12 \fi
    \number\hour:\ifnum\minute<100\fi\number\minute\the\amorpm}}
\edef\militarytime{\number\hour:\ifnum\minute<100\fi\number\minute}
\def\draftlabel#1{{\@bsphack\if@filesw {\let\thepage\relax
  \xdef\@gtempa{\write\@auxout{\string
    \newlabel{#1}{{\@currentlabel}{\thepage}}}}}\@gtempa
    \if@nobreak \ifvmode\nobreak\fi\fi\fi\@esphack}
     \gdef\@eqnlabel{#1}}
\def\@eqnlabel{}
\def\@vacuum{}
\def\draftmarginnote#1{\marginpar{\raggedright\scriptsize\tt#1}}
\def\draft{\oddsidemargin -.5truein
        \def\@oddfoot{\sl preliminary draft \hfil
        \rm\thepage\hfil\sl\today\quad\militarytime}
        \let\@evenfoot\@oddfoot \overfullrule 3pt
        \let\label=\draftlabel
        \let\marginnote=\draftmarginnote
   
\def\@eqnnum{(\theequation)\rlap{\kern\marginparsep\tt\@eqnlabel}%
\global\let\@eqnlabel\@vacuum}  }
\def\preprint{\twocolumn\sloppy\flushbottom\parindent 1em
        \leftmargini 2em\leftmarginv .5em\leftmarginvi .5em
        \oddsidemargin -.5in    \evensidemargin -.5in
        \columnsep 15mm \footheight 0pt
        \textwidth 250mmin      \topmargin  -.4in
        \headheight 12pt \topskip .4in
        \textheight 175mm
        \footskip 0pt
        
\def\@oddhead{\thepage\hfil\addtocounter{page}{1}\thepage}
        \let\@evenhead\@oddhead \def\@oddfoot{} \def\@evenfoot{} 
}
\def\titlepage{\@restonecolfalse\if@twocolumn\@restonecoltrue\onecolumn
     \else \newpage \fi \thispagestyle{empty}\c@page\z@
        \def\thefootnote{\fnsymbol{footnote}} }
\def\endtitlepage{\if@restonecol\twocolumn \else  \fi
        \def\thefootnote{\arabic{footnote}}
        \setcounter{footnote}{0}}  %\c@footnote\z@ }
\catcode`@=12
\relax
\def\ps@headings{\def\@oddfoot{}\def\@evenfoot{}
\def\@oddhead{\hbox{}\hfill
        \makebox[.5\textwidth]{\raggedright\ignorespaces --\thepage{}--
        \hfill }}
\def\@evenhead{\@oddhead}
\def\subsectionmark##1{\markboth{##1}{}}
}

\ps@headings

\relax

\def\firstpage#1#2#3#4#5#6{
\begin{document}
%%%%%%%%%%%%%%%%%%%%%%%%%%%%%%%%%%%%%%%%
%\draft
%%%%%%%%%%%%%%%%%%%%%%%%%%%%%%%%%%%%%%%%
\begin{titlepage}
\nopagebreak
\title{\begin{flushright}
        \vspace*{-1.8in}
        {\normalsize CERN--TH/97-179}\\[-8mm]
%       {\normalsize NUB--#1 #2}\\[-9mm]
   {\normalsize CPTH--S550.0797}\\[-8mm]
        {\normalsize hep-th/9707222}\\[4mm]
\end{flushright}
\vfill
{#3}}
\author{\large #4 \\[1.0cm] #5}
\maketitle
\vskip -7mm     
\nopagebreak 
\begin{abstract}
{\noindent #6}
\end{abstract}
\vfill
\begin{flushleft}
\rule{16.1cm}{0.2mm}\\[-3mm]
$^{\star}${\small Research supported in part by\vspace{-4mm}
the National Science Foundation under grant
PHY--96--02074, \linebreak and in part by the EEC \vspace{-4mm} 
under the TMR contract    
ERBFMRX-CT96-0090.}\\
$^{\dagger}${\small Laboratoire Propre du CNRS UPR A.0014.}\\
CERN--TH/97-179\\
July 1997
\end{flushleft}
\thispagestyle{empty}
\end{titlepage}}
%
%traces and imaginary parts
\def\Tr{\,{\rm Tr}\, }
\def\det{\,{\rm det}\, }
\def\Str{\,{\rm Str}\, }
\def\tr{\,{\rm tr}\, }
\def\Im{\,{\rm Im}\, }
\def\Re{\,{\rm Re}\, }
\def\im{\, {\rm Im}\, \tau}
\def\iS{S_2}
\def\iT{T_2}
\def\iU{U_2}
\def\ifd{\int_{\cal F}\frac{d^2\tau}{\t_2}}
% \def\ph{\vphantom}
% Greek letters
\def\a{\alpha}
\def\b{\beta}
\def\g{\gamma}
\def\d{\delta}
\def\eps{\epsilon}
\def\f{\phi}
\def\k{\kappa}
\def\l{\lambda}
\def\m{\mu}
\def\n{\nu}
\def\o{\omega}
\def\p{\pi}
\def\ps{\psi}
\def\Ps{\Psi}
\def\r{\rho}
\def\t{\tau}
\def\th{\vartheta}
\def\Th{\Theta}
\def\s{\sigma}
\def\x{\xi}
\def\ci{\chi}
\def\et{\eta}
\def\L{\Lambda}
\def\S{\Sigma}
\def\Ga{\Gamma}
\def\Fi{\Phi}
\def\O{\Omega}
\def\D{\Delta}

% barred variables
\def\thb{\bar{\th}}
\def\etb{\bar{\eta}}
\def\bz{\bar{z}}
\def\bet{\bar{\et}}
\def\fb{\bar{f}}
\def\bh{\bar{h}}
\def\bg{\bar{g}}
\def\bd{\bar{d}}
\def\kb{\bar{k}}
\def\lb{\bar{l}}
\def\bc{\bar{\g}}
\def\cb{\bar{c}}
\def\bu{\bar{u}}
\def\bw{\bar{w}}
\def\bv{\bar{v}}
\def\bq{\bar{q}}
 \def\bp{\bar{p}}
\def\br{\bar{r}}
\def\by{\bar{y}}
\def\bl{\bar{\l}}
\def\baa{\bar{a}}
\def\bb{\bar{b}}
\def\fb{\bar{f}}
\def\bi{\bar{\imath}}
\def\bj{\bar{\jmath}}
\def\bo{\overline{0}}
\def\tf{\bar{\phi}}
\def\bps{\bar{\psi}}
\def\bPs{\bar{\Psi}}
\def\bR{\overline{R}}
\def\bS{\overline{S}}
\def\bX{\overline{X}}
\def\bN{\overline{N}}
\def\bM{\overline{M}}
\def\bA{\overline{A}}
\def\bL{\overline{L}}
\def\bJ{\overline{J}}
\def\bV{\overline{V}}
\def\bQ{\overline{Q}}
\def\bI{\overline{I}}
\def\bP{\overline{P}}
\def\bQ{\overline{Q}}
\def\bG{\overline{G}}
\def\bE{\overline{E}}
\def\bF{\overline{F}}
\def\bT{\bar{T}}
\def\bU{\bar{U}}
\def\bC{\overline{C}}
\def\bX{\overline{X}}
\def\bY{\overline{Y}}
\def\bW{\overline{W}}
\def\bOmega{\overline{\Omega}}
\def\bLambda{\overline{\Lambda}}
% variables with tilde
\def\hh{\tilde h}
\def\tm{\tilde m}
\def\tu{\tilde{u}}
\def\tL{\widetilde{L}}
\def\tT{\widetilde{T}}
\def\tD{\widetilde{\Delta}}
\def\tX{\widetilde{X}}
\def\tPs{\widetilde{\Psi}}
\def\tbPs{\overline{\widetilde{\Psi}}}
% calligraphic variables
\def\sL{{\cal L}}
\def\J{{\cal J}}
\def\F{{\cal F}}
\def\cG{{\cal G}}
\def\P{{\cal P}}
\def\H{{\cal H}}
\def\cA{{\cal A}}
\def\cF{{\cal F}}
\def\G{{\cal G}}
\def\cD{{\cal D}}
\def\T{{\cal T}}
\def\C{{\cal C}}
\def\cO{{\cal O}}
\def\cQ{{\cal Q}}
\def\Y{{\cal Y}}
\def\B{{\cal B}}
\def\A{{\cal A}}
\def\N{{\cal N}}

% hatted variables
\def\Fh{\widehat{{\cal F}}}
\def\Lh{\widehat{L}}
\def\Mh{\widehat{M}}
\def\Bh{\widehat{{\cal B}}_H}
\def\Ch{\widehat{{\cal C}}_{g/h}}
\def\hG{\widehat{G}}
\def\ha{\hat{\a}}
\def\hb{\hat{\b}}
\def\C{{\cal C}}
\def\cO{{\cal O}}
\def\cQ{{\cal Q}}
\def\Y{{\cal Y}}
\def\B{{\cal B}}
\def\N{{\cal N}}
\def\simlt{\stackrel{<}{{}_\sim}}
\def\simgt{\stackrel{>}{{}_\sim}}
\newcommand{\dal}{\raisebox{0.085cm}
{\fbox{\rule{0cm}{0.07cm}\,}}}
\newcommand{\dt}{\partial_{\langle T\rangle}}
\newcommand{\dtbar}{\partial_{\langle\bar{T}\rangle}}
\newcommand{\al}{\alpha^{\prime}}
\newcommand{\mst}{M_{\scriptscriptstyle \!S}}
\newcommand{\mpl}{M_{\scriptscriptstyle \!P}}
\newcommand{\dv}{\int{\rm d}^4x\sqrt{g}}
\newcommand{\lv}{\left\langle}
\newcommand{\rv}{\right\rangle}
\newcommand{\ph}{\varphi}
\newcommand{\abar}{\bar{a}}
\newcommand{\sbar}{\,\bar{\! S}}
\newcommand{\xbar}{\,\bar{\! X}}
\newcommand{\fbar}{\,\bar{\! F}}
\newcommand{\zbar}{\bar{z}}
\newcommand{\dbar}{\,\bar{\!\partial}}
\newcommand{\tbar}{\bar{T}}
\newcommand{\taubar}{\bar{\tau}}
\newcommand{\ubar}{\bar{U}}
\newcommand{\tetabar}{\bar\Theta}
\newcommand{\etabar}{\bar\eta}
\newcommand{\qbar}{\bar q}
\newcommand{\ybar}{\bar{Y}}
\newcommand{\phb}{\bar{\varphi}}
\newcommand{\cm}{Commun.\ Math.\ Phys.~}
\newcommand{\prl}{Phys.\ Rev.\ Lett.~}
\newcommand{\pr}{Phys.\ Rev.\ D~}
\newcommand{\pl}{Phys.\ Lett.\ B~}
\newcommand{\ibar}{\bar{\imath}}
\newcommand{\jbar}{\bar{\jmath}}
\newcommand{\np}{Nucl.\ Phys.\ B~}
\newcommand{\Ftilde}{{\widetilde{\cal F}}}
\renewcommand{\L}{{\cal L}}
\newcommand{\I}{\widetilde{\cal I}}
\renewcommand{\Im}{\mbox{Im}}
\newcommand{\Zint}{{\mbox{\sf Z\hspace{-3.2mm} Z}}}
\newcommand{\Real}{{\mbox{I\hspace{-2.2mm} R}}}
\newcommand{\e}{e}
\newcommand{\be}{\begin{equation}}
\newcommand{\en}{\end{equation}}
\newcommand{\ba}{\begin{eqnarray}}
\newcommand{\ea}{\end{eqnarray}}
\newcommand{\ee}{\end{equation}}
\newcommand{\gsi}{\,\raisebox{-0.13cm}{$\stackrel{\textstyle
>}{\textstyle\sim}$}\,}
\newcommand{\lsi}{\,\raisebox{-0.13cm}{$\stackrel{\textstyle
<}{\textstyle\sim}$}\,}

\def\ft1{\tilde{F}_1}
\def\tm{\tilde m}\def\bul{$\bullet$}
\def\op{\oplus}
\def\oti{\otimes}
\def\pa{\partial}
\def\bpa{\bar{\partial}}
\def\na{\partial}
\def\ve{\vert}
\def\df{\delta \phi}
\def\bra{\langle}
\def\ket{\rangle}
\def\ra{\to}
\def\lra{\leftrightarrow}
\def\ti{\times}
\def\ss{\ar{a}{b}}
\def\rd{\rm d}
\def\bss{\ar{\ba}{\bb}}
\def\hg{\ar{h}{g}}
\def\w{\wedge}
\def\un{\underline}
\def\limit#1#2{\smash { \mathop{#1} \limits_{#2} }  }
\def\nn{\na\na}
\def\etab{ {\bar{\eta}} \raisebox{0.5ex}{$^{-24}$} (\bar{\tau}) }

\date{}
\firstpage{3155}{}
{\large\sc  Calculable $\;e^{-1/\lambda}\;$ Effects$^{\star}$ } 
%in $N=4$ Superstring Theory
{I. Antoniadis$^{\,a,b}$, B. Pioline$^{\,a,b}$
and T.R. Taylor$^{\,b,c}$}  %\\[-3mm] 
{\normalsize\sl
$^a$Centre de Physique Th{\'e}orique, Ecole Polytechnique,$^\dagger$
{}F-91128 Palaiseau, France\\[-3mm]
\normalsize\sl $^b$Theory Division, CERN, 1211 Geneva 23,
Switzerland\\[-3mm]
\normalsize\sl $^c$Department of Physics, Northeastern
University, Boston, MA 02115, U.S.A.}
{We identify and evaluate a class of physical amplitudes
in four-dimensional $N=4$ superstring theory,
which receive, in the weak coupling limit, 
contributions of order $\;e^{-1/\lambda}$, where $\lambda$
is the type II superstring coupling constant. They correspond to 
four-derivative $\Ftilde_1$ interaction terms involving the universal type II
dilaton supermultiplet. The exact result, obtained
by means of a one-loop computation in the dual heterotic theory
compactified on $\T^6$, is compared with the perturbation theory on the type II
side, and  the $e^{-1/\lambda}$ 
contributions are associated to non-perturbative 
effects of Euclidean solitons (D-branes) wrapped on $K3\times \T^2$.
The ten-dimensional decompactification limit on the type IIB side
validates the recent conjecture for 
the D-instanton--induced $R^4$ couplings.}

\section{Introduction}
The existence of non-perturbative string effects behaving 
like $e^{-1/\lambda}$ in the weak coupling limit $\lambda\to 0$ 
was predicted long ago 
by Shenker \cite{ss} by analyzing the large-order behavior of string
perturbation theory. Recent developments in superstring duality
give a new insight into the origin of non-perturbative effects, 
with membranes, five-branes and other solitons playing the central role
in understanding non-perturbative dynamics.
Although several possible sources of $e^{-1/\lambda}$ 
effects have already been identified \cite{bbs,ov,es,dix1,dix2},
their direct analysis still remains beyond the reach of computational 
techniques. However, duality itself provides a very powerful tool
for computing non-perturbative effects: the physical quantities
that can be determined exactly by using perturbative expansion in one theory
are often mapped to quantities receiving non-perturbative
contributions in the dual description
\cite{bfk}. By looking at the exact
results obtained in this way, we can try to identify the sources
of non-perturbative effects and hopefully learn how to compute them
directly.

An important example is provided by the series of higher-derivative
F-terms, $\F_g$ \cite{top}, which 
in four-dimensional $N=2$ supersymmetric compactifications of the heterotic
theory receive perturbative contributions and non-perturbative 
corrections; however, they originate entirely at genus $g$ on the type II side
\cite{agnt}. $\F_g$'s are holomorphic functions of $N=2$ vector moduli, up to a
holomorphic anomaly \cite{bcov}. In fact, $\F_1$ 
that determines the $R^2$ couplings \cite{agn}, appears already at the 
$N=4$ level
in type II compactifications on $K3\times \T^2$, as well as in heterotic
compactification on $\T^6$, where it receives only a tree-level and 
non-perturbative contributions.
Recently, Harvey and Moore
\cite{hm} obtained an exact result for $R^2$ couplings in $N=4$ theory
by performing a one-loop computation on the type IIA side and compared it
with the corresponding expression in the dual heterotic theory.
They identified terms that behave like
$e^{-1/\lambda^2}$ in the weak coupling limit $\lambda\to 0$ 
of the heterotic coupling constant and associated them with the effects
of gravitational instantons corresponding to
Euclidean five-branes wrapped on $\T^6$. In this case, analyticity and discrete
Peccei--Quinn symmetry forbid the presence of $e^{-1/\lambda}$ effects.

There exists another class of higher-derivative terms, $\Ftilde_g$ \cite{top},
which in the $N=2$ context corresponds to $(2g+2)$-derivative couplings
of the universal type II dilaton hypermultiplet $(S,Z)$, where
$S$ and $Z$ are the two $N=1$ chiral supermultiplet components. $S$ originates
from the Neveu-Schwarz--Neveu-Schwarz (NS-NS) sector and determines 
the type II coupling constant $\lambda$:
$\Im S\equiv e^{-2\varphi}=8\pi /\lambda^2$; $Z$ originates from the
Ramond--Ramond (R-R) sector. The function $\Ftilde_1$ determines the 
four-derivative coupling
\be
\I_{\rm II} ~=~ {\Ftilde_1 \over 2(\Im S)^2}\,
\big(\partial_{\mu}\partial_{\nu}\sbar\,\partial^{\mu} \partial^{\nu}
\sbar +\partial_{\mu}\partial_{\nu}S\,\partial^{\mu} \partial^{\nu}S\big)\ .
\label{ih}
\en
In type II perturbation theory, $\Ftilde_1$ 
is a harmonic function of the NS-NS hypermultiplet moduli 
(again up to a holomorphic anomaly), which is related to $\F_1$ by mirror
symmetry \cite{top}. Unlike $\F_1$, which is a purely one-loop quantity, 
$\Ftilde_1$ can also receive non-perturbative corrections.
These corrections break the 
mirror symmetry and allow a non-trivial dependence on
the R-R hypermultiplet components, violating perturbative holomorphicity.
For a type II model that admits a dual heterotic description, $\Ftilde_1$ 
can be computed exactly in the heterotic perturbation theory. 
There, it corresponds to a purely one-loop quantity because the heterotic
coupling constant belongs to a vector multiplet that decouples
from the hypermultiplet sector of the theory. Thus a one-loop heterotic
computation yields the exact result. 

Similarly to $\F_1$, $\Ftilde_1$ appears already for $N=4$ compactifications. 
In this work,
we obtain an exact answer for $\Ftilde_1$ in $N=4$ theory by means of
a one-loop computation in the heterotic theory compactified on
$\T^6$. We examine the limit that corresponds to the weak coupling regime of
the dual type II theory compactified on $K3\times \T^2$.
The result reproduces the perturbative type II expression \cite{top,sixguys},
and contains $e^{-1/\lambda}$ corrections that can be
associated to type II Euclidean solitons (D-branes) wrapped on $K3\times \T^2$.
In this case, such corrections are possible because the corresponding
couplings are not constrained to be holomorphic. 

On the type IIB side, $\Ftilde_1$ receives contributions from
the D-instantons already present in ten dimensions. Their
contribution can be isolated by decompactifying
the internal space $K3 \ti \T^2$. In $D=10$, $\Ftilde_1$ determines
the $R^2 (\na\na\ph_{10})^2$ coupling, where $\ph_{10}$ is the
ten-dimensional dilaton. This coupling is related 
by supersymmetry to $R^4$ couplings, for which an exact result
has been conjectured in Ref. \cite{dix1}.
As a by-product of our analysis,
starting from the exact four-dimensional expression for 
$\Ftilde_1$, we will prove the validity of this conjecture.

We first review some basic features of $N=4$ type II--heterotic
duality in $D=4$. Recall that in $D=6$, the type IIA theory on $K3$ and 
the heterotic theory on $\T^4$ are related by strong--weak coupling 
duality \cite{ht}. $N=4$, $D=4$ theories are obtained
by further compactification on $\T^2$. The global moduli space takes the form
\be \bigg[
\left. {Sl(2;\Real)\over U(1)}\right/ Sl(2;\Zint)\,\bigg] \;\times\;
\bigg[\left. {O(6,22;\Real)\over O(6)\times O(22)}\right/
O(6,22;\Zint)\,\bigg]\ .\label{mod}
\en
Viewed from the type IIA side, the first factor is associated
to the complexified K\"ahler modulus $T_{\rm II}$ of $\T^2$, while
the second factor, in the perturbative limit, decomposes into the product of the
NS-NS moduli:
\be
{O(6,22)\over O(6)\times O(22)}\supset
{O(4,20)\over O(4)\times O(20)}\times{SU(1,1)_S\over U(1)}
\times{SU(1,1)_U\over U(1)}\ ,
\label{dec}
\en
where the three cosets are associated with the 80 $K3$ $\sigma$-model moduli,
the IIA dilaton $S_{\rm II}$, and
the complex structure modulus $U_{\rm II}$ of $\T^2$, respectively.
This product is extended to the l.h.s. of Eq.\ (\ref{dec}) when
the Wilson lines $Y_{\rm II}$ of the 24 six-dimensional R-R gauge fields 
on $\T^2$ are taken into account.

{}From the heterotic point of view, the first factor of Eq.\ (\ref{mod}) 
is associated
with the heterotic dilaton $S_{\rm H}$, while the second factor is
associated to $T_{\rm H}$, $U_{\rm H}$, 
the 80 moduli of the $\Gamma_{4,20}$ Narain momentum lattice in
$D=6$, and the Wilson lines $Y_{\rm H}$ of the 24 six-dimensional 
gauge fields on $\T^2$. 
Type IIA--heterotic duality exchanges the dilaton $S$ of one side with
the $T$ modulus of the other side while identifying
$U$ and $Y$'s in the two theories. Our goal is to compute $\Ftilde_1$
in the heterotic theory compactified on $\T^6$
and to examine it in the limit of large K\"ahler modulus $T_{\rm H}$
(\ie\ the large $\T^2$ volume $\sqrt{G_{\rm H}}=\Im T_{\rm H}\to\infty$ limit),
which will take us to the weak coupling regime $\Im S_{\rm II}\to\infty$ on
the type IIA side. In the following, we will always consider
heterotic and type II fields in a well-defined context; 
from now on we therefore drop the subscripts H and II.

This paper is organized as follows. In Section 2, we determine the
one-loop perturbative $\Ftilde_1$ coupling in type 
IIA  theory on $K3\ti \T^2$.
In Section 3, we perform a one-loop computation in 
the dual heterotic string theory
compactified on $\T^6$,  and obtain the exact result for $\Ftilde_1$.
In Section 4, we examine the large two-torus limit of the exact result
and compare it with the perturbative result of Section 2.
We also obtain subleading contributions of order $e^{-1/\lambda}$.
We associate them in Section 5 with the non-perturbative
effects of Euclidean D-branes wrapping on $K3\ti \T^2$.
In Section 6, we examine the decompactification limit
of the dual type IIB theory and relate our result
to the ten-dimensional $R^4$ couplings.

\section{One-Loop $\Ftilde_1$ in Type II Theory}

A class of four-derivative interaction terms in type II superstring theory 
compactified on $K3 \ti \T^2$ have recently been considered 
in Ref.\ \cite{sixguys} at the one-loop level. In particular, all
terms quadratic in the dilaton $\ph$ and in the field-strength $H$ of
the NS-NS (two-index) antisymmetric tensor field have been obtained,
with the result written in the Einstein frame as:\vspace{-6mm}
\ba
{\I}&=&
 -\,{1\over 12} \Delta (U)\,\e^{-2\ph} \na_\m H_{\n\r\s}\na^\m H^{\n\r\s} 
-2 \Delta (U)\, \na_\m \na_\n \ph\, \na^\m \na^\n\!\ph \nonumber
\\
&&+\; {1\over 3}
\Theta(U)\, \eps^{\m\n\r\s} \e^{-2\ph} \na_\m \na_\a \ph\, \na^\a H^{\n\r\s}
+ 16~\e^{-2\ph} \epsilon^{\m\n\r\s} \pa_\m \pa_\a \ph\,H_{\n\r}^{\;\;\;\a} 
\,{\pa_\s \Re U \over \Im U}\ .
\label{eac}
\ea
In type IIA theory, the ``threshold'' functions are given by
\be
\label{thres}
\Delta(T,U)=-24~ \log(\Im U |\eta(U)|^4 )\ ,
\quad \, \quad
\Theta(T,U)=-24~ \Im \log( \eta^4(U) )\ ,
\ee
where $\eta$ is the Dedekind function. 
The result for type IIB is obtained by applying the mirror symmetry
$T \lra U$ in the above equations. 

The function $\Ftilde_1$ can be obtained from Eqs.\ (\ref{eac}) and 
(\ref{thres}) in the following way. First, by using the standard dualization
procedure, the antisymmetric tensor is replaced
by a scalar axion: 
$\eps_{\m\n\r\s}\na^\s\! a = e^{-2\ph}  H_{\m\n\r} + O(\na^3)$.
Then, the axion and the dilaton are combined into one
complex scalar $S\equiv a+ie^{-2\ph}$, which corresponds, 
from the $N=2$ point of view,
to the NS-NS component of the universal hypermultiplet.
After rewriting Eq.\ (\ref{eac}) in terms of the new variables,
we can extract the four-derivative coupling (\ref{ih}), with
\be
\label{ih11}
\Ftilde_1=-24 \log \left( \Im U |\eta(U)|^4 \right)\ .
\ee
The above expression is manifestly invariant under the perturbative
$Sl(2;\Zint)_{U}$ symmetry; however, it includes non-harmonic
terms that are due to integration of massless particles.
The full Wilsonian effective action $\widetilde{\cal I}_{\rm W}$
is obtained from Eq.\ (\ref{eac}) by subtracting
non-analytic $\Im\,U$ terms and reads:
\be
\label{deacS}
{\I}_{\rm W}=-24\log \eta^2(U)\;
{\partial_{\mu}\partial_{\nu}\sbar\,\partial^{\mu} \partial^{\nu}
\sbar\over 2(\Im S)^2} ~+~ {\rm c.c.} 
\ee

It is not accidental that the function appearing in Eq.\ (\ref{ih11})
is the same as $\F_1$ that determines the $R^2$ couplings
in $N=4$ type IIA theory at one loop, 
for which the corresponding Wilsonian action is \cite{hm}:
\be
\label{deacR}
{\cal I}_{\rm W}=-24\log \eta^2(T)\;
( R + i \widetilde{R} )^2 ~+~ {\rm c.c.} 
\ee
This relation is due to the
mirror symmetry, which exchanges $T$ with $U$ and maps 
the self-dual part of the Riemann tensor
$R + i \widetilde{R}$ to $(\na\na \sbar)/\Im S$ \cite{top}.
As mentioned before, the difference between the two functions
appears only at the non-perturbative level: unlike $\F_1$,
the function
$\Ftilde_1$  does receive non-perturbative corrections.
This has to be the case if the perturbative $Sl(2;\Zint)_U$ symmetry
is promoted to the full $SO(6,22;\Zint)$ duality, which also transforms the
type II dilaton.
We will discuss these corrections after deriving
the exact result from a one-loop computation
in the dual heterotic theory.

\section{Heterotic Computation of  $\Ftilde_1$}

We consider heterotic theory compactified on $\T^6$. We will extract
$\Ftilde_1$ from the interaction term 
\be
{\partial_{\phi_1}\partial_{\phi_2}\Ftilde_1\over 4(\Im T)^2}\;
\partial_{\mu}\phi_1\partial^{\mu}\phi_2\:
(\partial_{\nu}\tbar \partial^{\nu}\tbar
+\partial_{\nu} T \partial^{\nu} T)\ ,
\label{term}
\ee
which follows from Eq.\ (\ref{ih}) after replacing $\sbar$ by $\tbar$
and integrating by parts. 
This term can be determined from the four-point amplitude 
\be 
{\cal A}_{\f_1 \f_2} \equiv \langle \phi_1(p_1)
\phi_2(p_2)\tbar (p_3)\tbar (p_4)\rangle
\label{aterm}
\ee
involving two $\bT$ moduli of the two-torus
and two moduli $\f_1,\f_2$ of 
$O(6,22)\over O(6)\ti O(22)$. 
The corresponding kinematical structure is $(p_1p_2)(p_3p_4)$.

The $\T^2$ moduli are parametrized as usual by
\be
T=B_{12}+i\sqrt{G}\,\equiv T_1+iT_2\quad,\quad U=G_{12}/G_{22}+i\sqrt{G}/G_{22}
\,\equiv U_1+iU_2\ ,\label{tu}
\ee
where $G_{IJ}, ~I,J=1,2$, is the $\T^2$ metric
\be
\label{gij}
G_{IJ} = {\iT \over \iU} \left(\matrix{ 1 & U_1 \cr U_1 & |U|^2 }\right)
\ee
and $B_{12}$ is the NS
antisymmetric tensor field. The 
moduli vertex operators
are given in the zero-ghost picture by
\be
\label{vopm}
V_\f(p,\bz,z) =
v_{IJ}(\f) 
:[ \pa X^J (z,\bz) + i p_\m \ps^\m (z) \ps^J (z) ]~ \bpa X^I (z,\bz)
~\e^{i p_\m X^\m (\bz,z) }:\ ,
\ee
where $z$ is the position of the vertex operator
to be integrated on the world-sheet,
$X^\m,X^I$ are the spacetime and internal 
string coordinates, respectively, 
while $\ps^\m,\ps^I$ are their left-moving fermionic
superpartners;  
$v_{IJ}(\f)$
are the moduli wave function ``polarization tensors'', 
which for the two-torus moduli read:
\be
\label{vij}
v_{IJ}(\f) = \pa_\f (G_{IJ} - B_{IJ} )\ ,
\quad I,J=1,2.
\ee

The presence of (4,0) world-sheet supersymmetry 
requires four fermionic contractions on the
left-moving side, so that the corresponding part of the moduli vertex
operators (\ref{vopm}) can be restricted to its 
$p_\m\ps^\m\ps^J$ fermionic part. This already provides
four powers of momenta, and it is therefore sufficient for our purposes
to  set $p_i=0$ everywhere else.
The contractions of the internal fermions conserve
the $U(1)$ charge of the internal superconformal theory.
Under this symmetry all chiral moduli have charge $+1$, while
their complex conjugates have charge $-1$. 
This imposes that both $\f_1$ and $\f_2$ should be chiral.
At this point we restrict ourselves to $\f_1=\f_2=U$,
and shall discuss the general case later.

Then there are two possible contractions
of internal fermions
$\langle \ps(1) \ps(3) \rangle \langle \ps(2) \ps(4) \rangle
- (3 \leftrightarrow 4)$
and three 
contractions of the four spacetime fermions. 
After expressing the fermionic contractions
in terms of the Szeg\"o kernel, the  Riemann identity 
can be used to carry out the summation over (even) spin structures,
and we find the usual result that the
left-bosonic determinant is precisely cancelled and that the dependence
on the left-moving positions of the vertices disappears.
Taking into
account the symmetry under exchange of the two $\bT$ moduli,
one sees that only one of the above three contractions survives, yielding 
the desired $(p_1 p_2)(p_3 p_4)$ kinematical factor.

One is therefore
left with the correlator of the right-moving part
$\bpa X^I$  of the vertices
(\ref{vopm}). Any contraction of these will yield a total derivative
on the world-sheet, so that they can be safely restricted to their
zero-mode (classical) component $\bpa X^I_{\rm cl}$.
The integration on the positions of vertex operators thus
becomes trivial and yields a factor of $(\t_2)^4$.
After Poisson resummation,
which takes from Lagrangian to Hamiltonian representation, one finds
\be
\langle \dbar X^I\dbar X^J\dbar X^K\dbar X^L \rangle_{\rm cl}
=
\langle P_R^{I} P_R^{J} P_R^{K} P_R^{L} \rangle
- {\p\over \t_2}   G^{IJ} \langle P_R^{K}P_R^{L} \rangle
+ \left({\p\over \t_2}\right)^2 G^{IJ} G^{KL} + {\rm permutations,}
\ee
where $P_R$ are the right-moving momenta of 
the Narain lattice $\Gamma_{6,22}$.
The last two terms do not contribute to the amplitude
due to the first of the identities:
\ba
v_{IJ}(U) G^{IK} v_{KL}(\bT) &=& v_{IJ}(U) G^{IK} v_{KL}(U) = 0
\\
v_{IJ}(U) G^{JL} v_{KL}(\bT) &=& {2i\over \iT} v_{IK}(U)\ ,
\label{id2}
\ea
which follow from the definition (\ref{vij}). 
The amplitude can be further simplified by using the second
identity (\ref{id2}), and the final result reads:
\be
\A_{\phi_1\phi_2}\!=\frac{\p^2}{(\iT)^2}
\int_{\F} d^2 \tau \,\tau_2\!\!\!  \sum_{P_L,P_R \in\Gamma_{6,22}}
\left[ P_R^I v_{IJ}(\phi_1) P_R^J\right]
\left[ P_R^I v_{IJ}(\phi_2) P_R^J \right]\;
e^{i\pi\tau P_L^2} e^{-i\pi\taubar P_R^2}\,\etab\ .
\label{ampl}
\ee
In the above equation, the integration extends over the fundamental
domain of the Teichm\"uller parameter $\tau=\tau_1+i\tau_2$, and the sum
runs over the left- and right-moving momenta $P_L$ and $P_R$, respectively.
The Dedekind function factor
\be
\etab =\sum_{k\ge -1}C(k)\,
e^{-2\pi ik\tau_1}e^{-2\pi k\tau_2}\label{eta}
\ee
represents the contribution  of right-moving bosonic oscillators to the 
partition function, with the coefficients $C(k)$ counting the number 
of degenerate states at level $k$.

Let us now briefly discuss the generalization of Eq.\ (\ref{ampl})
to moduli $\f_1,\f_2$ other than $U$. Moduli of the $\Ga_{4,20}$
sublattice have wave functions $v(\f)$ orthogonal to $v(\bT)$ so that
the corresponding amplitude vanishes. This is also the
case for the Wilson line moduli of the four left-moving
gauge fields in $\Ga_{4,20}$. On the other hand, chiral Wilson lines
$W^R_i=Y_{1i}+U Y_{2i}$
of the 20 right-moving gauge fields yield non-zero contractions,
and Eq.\ (\ref{ampl}) is still valid in that case, with the appropriate
wave functions $v(W_i)$. From now on, we will
consider the case where $\f_1,\f_2$ denote $U$ or $W^R_i$ moduli.

\section{Large-Volume Limit and Comparison with Type II Theory}

In order to study the large $\iT$ behavior of the amplitude
(\ref{ampl}), we need to be more specific about the lattice
decomposition of $\Gamma_{6,22}$ into $\Gamma_{2,2}\oplus\Gamma_{4,20}$.
The momenta of the $\Gamma_{4,20}$ lattice are parametrized by the
integer charges $q^i$, $i=1,\dots, 24$, with the left- and right-moving norms
\be
P_L^2={1\over 2}q^t (\Mh+\Lh) q \quad ,\quad P_R^2={1\over 2}q^t (\Mh-\Lh) q\ ,
\label{a1}
\ee
where $\Mh$ is the symmetric $24\times 24$ matrix of moduli parametrizing 
the coset space\linebreak
$O(4,20)\over O(4)\times O(20)$; $\Mh$ is orthogonal 
with respect to the signature (4,20) metric $\Lh$: $\Mh^{-1}=\Lh \Mh \Lh$.

The $\Gamma_{6,22}$ lattice is now constructed by
supplementing the 24 $\Gamma_{4,20}$ charges
$q_i$ with two momentum numbers $m_I$ and two winding numbers
$n^I$ on $\Gamma_{2,2}$, and including $2\times 24$ Wilson lines 
$Y_{Ii}$, which boost
a general $\Gamma_{6,22}$ lattice back to $\Gamma_{2,2} \oplus \Gamma_{4,20}$.
Following Ref.\ \cite{KR}, we write the left- and right-moving momentum
norms as
\be
\label{a2}
P_L^2= {1\over 2}p_L^t G^{-1} p_L + {1\over 2} Q^t (\Mh+\Lh) Q
\quad ,\quad
P_R^2= {1\over 2}p_R^t G^{-1} p_R + {1\over 2} Q^t (\Mh-\Lh) Q\ ,
\label{aa11}
\ee
where
\ba
p_{L;I}&=&
m_I+Y_{Ik}q^k-{1\over2}Y_{Ii}\Lh^{ij}Y_{Jj}n^J+B_{IJ}n^J+G_{IJ}n^J\ ,
\nonumber\\
p_{R;I}&=&
m_I+Y_{Ik}q^k-{1\over2}Y_{Ii}\Lh^{ij}Y_{Jj}n^J+B_{IJ}n^J-G_{IJ}n^J\ ,
\label{a3}
\ea
are the boosted momentum components on the $\T^2$ torus,
and the charges are given by
\be
Q^i=q^i-\Lh^{ij}Y_{Ij}n^J\ . \label{a31}
\ee

In terms of the 28 charges $\cQ\equiv (m_I,n^I,q^i)$, Eq.\ (\ref{aa11}) can be
recast into the standard form:
\be
\label{a4}
P_L^2= {1\over 2}
\cQ^t (M +L) \cQ \quad ,\quad P_R^2= {1\over 2}\cQ^t (M -L) \cQ\ , 
\ee
where now $L$ is the signature (6,22) metric
\be
L=\left(\matrix{0&{\bf 1}_{2}&0\cr
{\bf 1}_{2}&0&0\cr 0&0&\Lh\cr}\right)
\label{a5}
\ee
and $M$ is the symmetric $O(6,22)\over O(6)\times O(22)$ moduli matrix:
\be
M=\left(\matrix{
G^{-1}     & G^{-1}C             &G^{-1}Y^{t}\cr
C^{t}G^{-1}&G+C^{t}G^{-1}C+Y^{t}\Mh^{-1}Y&C^{t}G^{-1}Y^{t}-Y^{t}\Lh\Mh\cr
YG^{-1}    &YG^{-1}C-\Mh\Lh Y  &\Mh+YG^{-1}Y^{t}\cr}\right),
\label{a6}
\ee
with
\be
C_{IJ}=B_{IJ}-{1\over 2}~Y_{Ii}\Lh^{ij} Y_{Jj}\ .
\label{a7}
\ee
The amplitude (\ref{ampl}) can be now written as
\be
\A_{\phi_1\phi_2}\!=\frac{\p^2}{(\iT)^2}
\int_{\F} d^2 \tau \,\tau_2
\!\!\!  \sum_{\cQ\in\Gamma_{6,22}}
\left[ P_R^I v_{IJ}(\phi_1) P_R^J\right]
\left[ P_R^I v_{IJ}(\phi_2) P_R^J \right]\;
e^{-\pi\tau_2\cQ^t M\cQ +i\pi\tau_1 \cQ^t L\cQ}\,\etab\ .
\label{ampl2}
\ee

In the large $\iT$ limit, with the other moduli being fixed, the
$\T^2$ metric $G_{IJ}$ (\ref{gij}) scales uniformly as $\iT$, so that 
$\cQ^t M\cQ \to n^t G n + O(1)$. The contributions
of states with non-vanishing winding numbers $n^I$ to the amplitude
(\ref{ampl2}) are therefore exponentially suppressed as
$\e^{-\iT}$. 
In the first approximation,
we can neglect all winding states; from Eq.\ (\ref{a3}) it follows that the 
remaining ones satisfy $P_{L;I}=P_{R;I}$. Then we can use the identity
\be
\sum_{P_L,P_R \in\Gamma_{6,22}}
P_L^I v_{IJ}(\phi) P_R^J
e^{i\pi\tau P_L^2} e^{-i\pi\taubar P_R^2}=\frac{1}{\pi\tau_2}\partial_{\phi}
\sum_{P_L,P_R \in\Gamma_{6,22}}
e^{i\pi\tau P_L^2} e^{-i\pi\taubar P_R^2}
\ee
to rewrite Eq.\ (\ref{ampl2}) as
\be
\A_{\phi_1\phi_2} ~\approx~ \frac{1}{(\iT)^2}D_{\phi_1}D_{\phi_2}
\int_{\F} {d^2 \tau \over\tau_2}  \sum_{m_I,q^i}
e^{-\pi\tau_2\cQ^t M\cQ +i\pi\tau_1 \cQ^t L\cQ}\,\etab\ ,
\label{ampl3}
\ee
where now
\be
\cQ^t M \cQ = m^t G^{-1} m + 2 m^t G^{-1} Y^{t} q + 
q^t (\Mh + Y G^{-1} Y^t ) q\ ,
\ee
\be
\cQ^t L \cQ = q^t \Lh q\ .
\ee
In Eq.\ (\ref{ampl3}), $D_\f$ denotes the K\"ahler 
covariant derivative, which for $\f=U$
coincides with the usual $Sl(2;\Zint)$ modular 
covariant derivative. In this case,
$D_{U} D_{U} = \left( \pa_U - {i\over \iU} \right) \pa_U  $.
The covariantizing term is
due to one-particle reducible diagrams, involving massless states
that have to be subtracted in order to obtain the (1PI) effective action. 
We recall that Eq.\ (\ref{ampl3}) only holds for $U$ and $W^R_i$ moduli,
whereas ${\cal A}_{\f_1\f_2}$ vanishes for $\Ga_{4,20}$ moduli
and for remaining Wilson lines. Integrability therefore occurs only
in the large $T_2$ limit, when the winding states are suppressed, and for
the $U$, $W^R_i$ moduli only. For the latter,
the scattering amplitude can be integrated to:
\be
\Ftilde_1 =
\int_{\F} {d^2 \tau \over\tau_2}  \sum_{m_I,q^i}
e^{-\pi\tau_2\cQ^t M\cQ +i\pi\tau_1 \cQ^t L\cQ}\,\etab\ ,
\label{ft1h}
\ee 
where the other moduli are treated
as constant background fields. Note that the above expression
displays $O(4,20;\Zint)\ti Sl(2;\Zint)_{U}$ invariance only,
since the $\Ftilde_1$ coupling singles out the $\tbar$ modulus.

After performing a Poisson resummation $m_I\to\tm^I$, we can rewrite
\be
\label{ft11h}
\sum_{m_I} e^{-\p\t_2 \cQ^t M \cQ~+~i\p\t_1 \cQ^t L \cQ }
=
{\iT\over \t_2} \sum_{\tm^I}
e^{-{\p\over\t_2}\tm^t G \tm~-~\p\t_2 q^t\Mh q
~-~2\p i \tm^t Y^t q ~+~\p i\t_1 q^t\Lh q }
\ee
The term with $\tm^I=0$ gives back the partition function
of $\Ga_{4,20}$. This term depends neither on $U$ nor
on the Wilson lines, and thus does not contribute to Eq.\
(\ref{ampl3}). We shall therefore ignore it
in the following.\footnote{However, this term will be important for
our discussion of the $\langle TT\tbar\tbar\rangle$ amplitude
${\cal A}_{TT}$ in Section 6.}
On the other hand, terms with $(\tm^1,\tm^2)\ne (0,0)$
contribute only in the region of large $\t_2$, for which 
$\t_1 \in [-{1\over 2},{1\over 2}]$. The integral over $\t_1$
imposes the level matching condition $k=q^t \Lh q/2$, so that
\be
\Ftilde_1 ~=~ \iT
\int_0^{\infty} 
{d\tau_2 \over(\tau_2)^2}  
{\sum_{\tm^I,q^i}}'
C\left({q^t \Lh q \over 2}\right)
\e^{-{\p\over \t_2}\tm^t G \tm~-~\p\t_2 q^t(\Mh+\Lh) q
~-~2\p i \tm^t Y^t q}
+O(e^{-\iT})\ ,
\label{ft2h}
\ee
where the coefficients $C(k)$ are defined in 
Eq.\ (\ref{eta}) and
the primed sum runs over $(\tm^1,\tm^2)$ $\ne (0,0)$
and unrestricted $q$'s. We set the lower bound of the integral to 0,
thereby neglecting terms of the same order as 
the winding-state contributions.

States with non-zero $q^i$ charges have
strictly positive $q^t (\Mh+\Lh) q$; 
they are therefore exponentially
suppressed with respect to the $q^i=0$ neutral states.
Hence the dominant contribution to $\Ftilde_1$ therefore picks
states with $C(q^t \Lh q/2 =0)=24$:
\be 
\label{ampl4}
\Ftilde_1 ~=~24~\iT
\int_{0}^{\infty} {d\t_2 \over (\t_2)^2} {\sum_{\tm^1,\tm^2}}'
e^{-\p{\iT\over \iU\t_2} |\tm^1 + U \tm^2|^2 }~+~\delta\Ftilde_1
+ O(e^{-\iT})\ ,
\ee
where $\delta\Ftilde_1$ denotes the contribution of the
$q^i\ne 0$ charged states to the integral of Eq.\ (\ref{ft2h}).
The integral in Eq.\ (\ref{ampl4})
was evaluated before in Ref.\ \cite{dkl} with the result:
\be
\label{dkldeg}
\int_0^{\infty} {d \t_2 \over( \t_2)^2} \iT {\sum_{\tm^1,\tm^2}\hskip -1mm}'
e^{-\p{\iT\over \iU\t_2} |\tm^1 + U \tm^2|^2 }
= - 4 \Re \log \eta(U) - \log ( \iU\iT ) \ ,
\ee
up to a moduli-independent, logarithmic divergence.
This divergence, together with the $\log\iT$ term,
do not contribute to the amplitude (\ref{ampl3}) and can be ignored.
We therefore obtain, in the leading $\iT\to\infty$ approximation,
\be
\label{h1ii}
\Ftilde_1 = -24 \log \left(U_2 |\eta(U)|^4 \right) + \dots
\ee
in agreement with the one-loop type IIA result (\ref{ih11}).
 
We now turn to the next-to-leading corrections $\delta\Ftilde_1$,
due to subleading terms in Eq.\ (\ref{ft2h}).
They originate from states with non-vanishing $q^i$ charges,
but still satisfying the level matching condition $k = q^t \Lh q /2$.
The main contribution comes from the saddle point
$\t_2^* = \sqrt{ \tm^t G \tm / q^t (\Mh + \Lh) q} \sim \sqrt{\iT}$.
The integral can be expressed
in terms of the Bessel function $K_{1}$, with the result:
\be
\label{amplK}
\delta\Ftilde_1 = 2 \iT
{\sum_{\tm^I,q^i}}''
C\left({q^t \Lh q \over 2}\right)
\left[ {q^t (\Mh + \Lh) q \over \tm^t G \tm} \right]^{1/2}
K_1 \left( 2\p \sqrt{ \tm^t G \tm \cdot 
{q^t (\Mh+\Lh) q \over 2}}\right)
\e^{-~2\p i \tm^t Y^t q}\ , 
\ee
where the double-primed sum runs over $(\tm^1,\tm^2)\ne(0,0), ~q^i \ne 0$. 
In the saddle-point approximation, 
$K_1(z)=\sqrt{\p\over 2z} e^{-z} [1 + O(1/z)]$, we obtain:
\be
\label{ampl5}
\delta \Ftilde_1 ~\approx~ 2 \iT
{\sum_{\tm^I,q^i}}''
C\left({q^t \Lh q \over 2}\right)
\left[ {q^t (\Mh + \Lh) q \over 2 
\left(\tm^t G \tm \right)^3} \right]^{1/4}
\e^{-2\p \sqrt{ \tm^t G \tm \cdot {q^t (\Mh+\Lh) q \over 2}} 
~-~2\p i \tm^t Y^t q} 
\ee
up to power-suppressed $O( T_2^{-{n\over 2}} e^{-\sqrt{\iT}})$ terms.

The final result, rewritten in terms of type IIA variables,
becomes
\ba
\Ftilde_1 \!&=&\!\! -24 \log \left( \iU |\eta(U)|^4 \right) ~+~
2 S_2^{1/4}\!\!
{\sum_{\tm^I,q^i \atop q^t\Lh q\ge -2}\hskip -2mm}''
C\left({q^t \Lh q \over 2}\right)
\left[{ q^t (\Mh + \Lh) q \over 2}\right]^{1/4}
\left[ {|\tm^1 + U \tm^2|^2\over \iU}  \right]^{-3/4}
\nonumber
\\
\label{ampl6}
&&\hspace{5.9cm}\ti\,\e^{-2\p \sqrt{\iS ~{|\tm^1 + U \tm^2|^2\over \iU}
 \cdot {q^t (\Mh+\Lh) q \over 2}} ~-2\p i \tm^t Y^t q} + \dots ,
\ea
where the neglected terms include the subleading saddle-point
contributions $O( S_2^{-{n\over 2}} e^{-\sqrt{\iS}})$ as well
as the contributions of the heterotic winding states  
$O( e^{-\iS})$. The above equation explicitly shows the violation
of harmonicity already at order $O(e^{\sqrt{S_2}})$, 
by non-perturbative type IIA corrections.

\section{$e^{-1/\lambda}$ Effects and D-Brane Instantons in Type IIA Theory}

As we argued in the Introduction, the heterotic one-loop
amplitude (\ref{ampl2}) gives the exact result for the
$\Ftilde_1$ coupling (\ref{term}). Viewed from the type IIA
side, the result contains, in addition to the one-loop contribution
(\ref{ih11}), a sum of non-perturbative terms $\delta \Ftilde_1$,
Eq.\ (\ref{amplK}),
of order $e^{-\sqrt{\iS}} \sim e^{-1/\lambda}$, as well
as subleading contributions of order
$e^{-\iS} \sim e^{-1/\lambda^2}$.
In the following, we will trace the origin of these terms
to the effects of Euclidean solitons of type IIA superstring.

All ten-dimensional superstrings have in common a
NS 5-brane \cite{five}, a 5+1 extended object magnetically charged
under the NS antisymmetric two-form.
In type IIA, the 5-brane is described by a chiral (5+1)-dimensional
world-volume theory with $(2,0)$ supersymmetry, 
while the IIB and heterotic
5-branes are non-chiral. The presence
of additional R-R gauge potentials 
in type II superstrings  allows the existence of
further soliton configurations charged under these gauge fields,
known as Dirichlet D$p$-branes \cite{P}. 
Type IIA superstring possesses
a one-form and a three-form R-R gauge potential, and consequently 
D$p$-branes with $p=0,2,4,6,8$. 
These solitons preserve one-half
of the ten-dimensional supersymmetry, while they break
the other half, generating fermionic zero-modes. 
Upon compactification, branes 
with a Euclidean world-volume can be wrapped  
on the compact manifold to yield instanton configurations in lower
dimensions.  These configurations can correct the low-energy 
effective action, but owing to supersymmetry
only solitons with a definite number
of fermionic zero modes can correct terms with a given number of
derivatives. 

In the case at hand, the coupling 
$\Ftilde_1 (\na\na\sbar)^2$
is related by supersymmetry to an eight-fermion coupling, so that
only instantons with at most eight zero-modes, breaking 
one-half of  $N=4$ supersymmetry, can contribute. These instantons
can be obtained by wrapping D0-, D2-, D4-branes or the NS 5-brane
on $K3\ti \T^2$, but only on minimal cycles of the compactification
manifold, under penalty of further breaking supersymmetry and thereby
generating extra zero-modes. 
The D-branes therefore have to wrap
one of the 24 even cycles of $K3$ times a circle $S_1$ in $\T^2$, while
the NS 5-brane has to wrap $K3\ti \T^2$. The latter ones are known
to generate $e^{-1/\lambda^2}$ effects for $R^2$ couplings
in four-dimensional toroidal compactification  of the heterotic 
string \cite{hm}. In our case, the $O(e^{-1/\lambda^2})$ corrections 
due to heterotic winding states are similarly generated by the 
chiral IIA 5-brane wrapping on $K3 \ti \T^2$. 
Here we will show that D-branes wrapping on $K3 \ti S_1$
have the required classical action to account for
the $e^{-1/\lambda}$ effects found in $\Ftilde_1$.

In order to evaluate this classical action, we note that the four-dimensional 
D-instantons can be constructed in two steps, first
by wrapping $p$ dimensions of the $(p+1)$-dimensional brane world-volume
on an even cycle of $K3$, thereby obtaining a point-like
soliton configuration
in six dimensions, and subsequently wrapping the soliton world-line
on one cycle of $\T^2$. The classical action can therefore be written
as the product of the length $\ell$ of the world-line times the 
mass $\cal{M}$  of the soliton. A minimal world-line, winding
$\tm^1$ times on the first circle and $\tm^2$ times on the second circle
of the torus, has length
\be
\label{len}
\ell= \sqrt{ \iT  {|\tm^1 + U \tm^2|^2 \over \iU} }
\ee
in string units.
On the other hand,
the mass formula for six-dimensional BPS states can be easily obtained
from the heterotic side. BPS states in the perturbative heterotic
spectrum have no oscillator excitations ($N_L=0$) on the (supersymmetric) 
left side, while the level-matching condition imposes 
$N_R-1={1\over 2}(P_L^2-P_R^2)= {1\over 2} q^t \Lh q$
on the right side. Their mass squared is therefore
\be
{\cal M}^2_{\rm het}= {1\over 2}P_L^2 + {1\over 2} P_R^2 + N_L + N_R -1
= P_L^2 = {1\over 2} q^t (\Mh + \Lh ) q\ ,
\ee
in heterotic string units.
Using heterotic--type IIA strong--weak coupling duality 
in six dimensions, this result translates
into
\be
\label{mass2}
{\cal M}^2= e^{-2\varphi_6}{ q^t (\Mh + \Lh ) q \over 2}
= {\iS \over \iT} {q^t (\Mh + \Lh ) q  \over 2}
\ee
in type IIA string units, with $e^{-2\varphi_6} = \iS /\iT$,
where $\varphi_6$ is the six-dimensional type IIA dilaton.
This formula gives the mass of BPS states in $D=6$
in terms of their charges $q^i$ under the 24 R-R
gauge fields. States with non-zero $q^i$ are non-perturbative
from the type IIA point of view, since their masses scale as
${\cal M}\sim 1/\lambda$, a characteristic feature of D-branes.
As a check on Eq.\ (\ref{mass2}), recall that
the mass of a D0-brane is  (in string units)
$e^{-\varphi_{10}}$, 
where $\varphi_{10}$ is the ten-dimensional dilaton. Similarly,
the mass of a D4-brane
wrapped on $K3$ is $e^{-\varphi_{10}}V$, where $V$ is the $K3$ volume.
Using $e^{-2\varphi_{6}}=V e^{-2\varphi_{10}}$,
we find ${\cal M}^2({\rm D}0) = e^{-2\varphi_6} / V$,
${\cal M}^2({\rm D}4) = e^{-2\varphi_6} V$.
This agrees with the decomposition of the moduli of $K3$,
\be
\label{deck3}
{O(4,20)\over O(4)\ti O(20)} \supset {O(3,19)\over O(3)\ti O(19)} 
\ti \Real^+_{V}\ ,
\ee
analogous to Eq.\ (\ref{a6}). In this decomposition,
$\Real^+_{V}$ parametrizes the volume of $K3$, whereas
${O(3,19)\over O(3)\ti O(19)}$ parametrizes the 
unit volume Ricci-flat metric \cite{asp}.

Putting Eqs.\ (\ref{len}) and (\ref{mass2}) together, we obtain the 
classical Euclidean action of a four-dimensional D-instanton:
\be
S_{\rm cl} = \ell \cdot {\cal M} = \sqrt{ 
\iS { |\tm^1 + U \tm^2|^2 \over \iU}\cdot
{q^t (\Mh + \Lh) q \over 2} }\ .
\ee
These are precisely the exponential weights occurring
in the expansion (\ref{ampl6}).

It now remains to interpret the phase factor $e^{-2\p i \tm^t Y^t q}$.
The charges $q^i$ of D-branes wrapped on a cycle $\gamma$ of $K3$
are determined by the expansion of $\gamma$
in terms of a basis $\gamma_i$ of the integer homology of $K3$:
$\gamma = \sum_{i} q^i \gamma_i $. Note that the (4,20) metric
$\Lh= \int_{K3} \gamma_i \w \gamma_j$ 
is the intersection matrix of the homology basis $\gamma_i$.
On the other hand, the Wilson lines $Y_{Ii}$
correspond to the $\T^2$ components of the six-dimensional 
gauge fields $A_i$, obtained from the reduction
of the ten-dimensional R-R gauge
potential $A_{RR}$ on $K3$: $A_{RR} = \sum_{i} A_i \w \gamma_i$
($A_{RR}$ denotes the formal sum of the R-R gauge potentials,
and similarly $\gamma$ is a formal sum of integer homology cycles of $K3$).
The phase $\tm^t Y^t q$ can then be
expressed as:
\be
\tm^t Y^t q =
\int_{S_1} q^i A_i =
\int_{S_1\ti K3} \gamma \w \gamma_i \w A_i =
\int_{S_1\ti K3} \gamma \w A_{RR} =
\int_{S_1\ti \gamma} A_{RR}\ . 
\ee
Hence the phase originates from the minimal D-brane--gauge potential
coupling.

A few remarks on the prefactors occurring in Eq.\ (\ref{ampl6})
are in order. The Dedekind function coefficient 
$C(q^t \Lh q /2)$
can be interpreted as the degeneracy of BPS states with charges
$q^i$ (this holds for the perturbative heterotic BPS spectrum,
and should remain valid for the complete
spectrum also). On the other hand, in the type IIA description, 
$\sigma=q^t\Lh q$ gives the 
the self-intersection number of the even cycle on which the 
D0-, D2- and D4-branes wrap. It must therefore be the case that
$C(\sigma/2)$ gives the number of supersymmetric cycles with
self-intersection $\sigma$ in $K3$, 
as has already been argued in Ref.\ \cite{bsv}. This could
in principle be proved from the interpretation of the  
modular form $1/\eta^{24}$ as the elliptic genus of $K3$ \cite{kawai}.
The restrictions $\sigma=0~ ({\rm mod}\  2)$ and $\sigma \ge -2$
are easily understood in the case of D2-branes wrapping a 2-cycle of
$K3$, since for a genus $g$ algebraic curve,
the self-intersection equals the Euler number $2g-2$.
That in turn confirms that BPS states are only obtained from 
such algebraic cycles, and not, say, from 2-cycles with boundaries.
It would be interesting to geometrically
understand the restriction $\sigma\ge -2$
for the case of bound states of D0-, D2- and D4-branes, where
the D0- and D2-branes now appear as Yang-Mills instantons and fluxes
on the $K3 \times \Real$ world-volume of the D4-brane \cite{pk}. 
At present we do not know of any obvious explanation
of the other prefactors, but we hope that they can provide a clue
towards a consistent treatment of D-brane instantons.

Finally, it is interesting to examine the large $\T^2$ decompactification
limit of the type IIA coupling $\Ftilde_1$ in Eq.\ (\ref{ampl6}).
The four-dimensional dilaton is related to the six-dimensional one through
$\iS=\iT e^{-2\ph_6}$. Sending $\iT\to\infty$, while keeping $\ph_6$
and the other moduli fixed, eliminates all the $O(e^{-1/\lambda})$
corrections from $\Ftilde_1$. This is in agreement with the fact that
the (odd-dimensional) world-volume of the type IIA D-branes cannot
be wrapped on the (even-dimensional) cycles of $K3$. Thus, there are
no D-brane instantons corrections to the type IIA theory
compactified on $K3$ (nor NS 5-brane corrections). 
The type IIA one-loop contribution
(\ref{h1ii}) is furthermore suppressed by a power of the volume $\iT$
of the two-torus and disappears in $D=6$. This is similar to the
vanishing of the one-loop $R^2$ corrections in type IIB theory 
on $K3$ \cite{sixguys}. We conclude that there are
no $(\pa\pa\ph_6)^2$ interactions in $D=6$ type IIA theory. 
As we show in the next Section,
this is not the case in type IIB theory.

\section{$D=10$ Decompactification Limit and Type IIB D-instantons}
We will now discuss the exact result (\ref{ampl}) in terms of
the type IIB perturbation theory. The type IIB theory is related 
to type IIA theory by $T$-duality that exchanges $T$ and $U$.
The type IIB perturbation theory is still organized 
as a large $S_{\rm IIB}=T_{\rm H}$ expansion. Therefore the exact result
can be rewritten as the weak type IIB coupling expansion:
\ba
\label{amplB}
\Ftilde_1 &=& -24 \log\left(\iT |\eta(T)|^4 \right)
~+~2 \iS
{\sum_{\tm^I,q^i}}''
C\left({q^t \Lh q \over 2}\right)
\left[ 
{T_2\over S_2} {q^t (\Mh + \Lh) q \over |\tm^1 + T \tm^2|^2} \right]^{1/2}
\cr
&&\ti~K_1 \left( 2\p \sqrt{ S_2  {|\tm^1 + T \tm^2|^2\over \iT} \cdot 
{q^t (\Mh+\Lh) q \over 2}}~\right)
\e^{-2\p i \tm^t Y^t q} + O(e^{-1/\lambda^2})
\ea
obtained from Eq.\ (\ref{ampl6}) by substituting $U_{\rm IIA}$ by $T_{\rm IIB}$.
The above result still exhibits $O(e^{-\sqrt{S_2}})$ corrections
that should similarly be interpreted in terms of type IIB Euclidean
D-instantons. Those can be obtained as D1-, D3-, D5-branes now wrapping
around even cycles of $K3\ti \T^2$, as well as D($-1$)-instantons,
which are already present in the ten-dimensional IIB theory.
Here we will focus on the latter contributions, which can be isolated
in the decompactification limit of the internal six-dimensional
manifold: $\iT\to\infty, V \to \infty$.

Taking first the large $\iT$ limit, we can restrict the summation
in Eq.\ (\ref{amplB}) to $\tm^2=0,\ \tm^1\ne 0$. In this limit,
the contribution of the D5-brane and of the D1-strings wrapping
on $\T^2$ disappears (together with that of the NS 5-brane)
and we obtain the following result for the six-dimensional coupling
$\Ftilde_1^{(6)} (\pa\pa\varphi_6)^2$ (in string 
units):\footnote{This result should be exact
in the six-dimensional limit without further $O(e^{-1/\lambda^2})$ 
corrections, because of the absence of NS 5-brane instantons.}
\be
\label{amplB6}
\Ftilde_1^{(6)} \approx 8\p 
~+~2 e^{-\ph_6}\! 
{\sum_{\tm^1\ne 0 \atop q^i\ne 0}}
C\!\left({q^t \Lh q \over 2}\right)
{ \sqrt{ q^t (\Mh + \Lh) q } \over |\tm^1|}
~K_1 \left( 2\p |\tm^1| 
\sqrt{q^t (\Mh+\Lh) q \over 2} e^{-\ph_6} \right)
\e^{-2\p i \tm^1 Y_{1i} q^i} 
\ee
where we extracted the $\T^2$ volume.
Note that the above equation involves only one-half of the
four-dimensional ``Wilson line'' $Y$ fields. This is consistent
with the fact that in type IIB, half of the $Y$'s correspond
to the R-R fluxes on $K3$, while the other half describes the
internal $B_{12}$ components of the 24 six-dimensional
R-R antisymmetric tensors. 

The large $V$ limit can be studied by considering the decomposition 
(\ref{deck3}) of the ${O(4,20)\over O(4)\ti O(20)}$ $K3$ 
moduli space. This corresponds to
a block decomposition of the matrix $\Mh$, similar to Eq.\ (\ref{a6}),
in which the first diagonal block represents the zero-cycle charge with
$\Mh_{00}=1/V$, the second represents the four-cycle charge with
$\Mh_{44}=V$, whereas the last $22\ti 22$ diagonal block describes
the 22 two-cycle charges and is independent of the volume $V$.
Therefore only D$(-1)$-instantons survive in the large $K3$ limit.
In this case, $q^t \Lh q =0$,  and Eq.\ (\ref{amplB6}) becomes
(in string units):
\be
\label{amplB7}
\Ftilde_1^{(10)} =
8\p + 48 e^{-\ph_{10}} 
{\sum_{\tm\ne 0, q\ne 0}}
\left| {q \over \tm} \right|
~K_1 \left( 2\p |q \tm| e^{-\ph_{10}} \right)
\e^{-2\p i \tm q a} \ ,
\ee
where we used $e^{-2\ph_{6}} = Ve^{-2\ph_{10}}$ and denoted
by $a$ the remaining ``Wilson line'' modulus $Y_{1,i=0}$ 
that corresponds to the ten-dimensional R-R scalar. Together
with the dilaton, they form a complex scalar
$\rho=a+ i e^{-\ph_{10}}$ transforming as a modulus
of the  $Sl(2;\Zint)_{\rho}$ type IIB symmetry \cite{ht}.\footnote{In four 
dimensions,
this symmetry is part of the full $O(6,22;\Zint)$ symmetry.}

The above expression (\ref{amplB7}) coincides with the ten-dimensional
$f(\rho,\bar{\rho})~R^4$ coupling conjectured in Ref. \cite{dix1}:
\be
\label{gg}
{12\over \p}  f(\rho,\bar{\rho})={24 \zeta(3)\over\p} (\rho_2)^2
+ 8\p + 48 \rho_2 \sum_{q\ne 0,\tm\ne0 }
\left|{q\over \tm} \right|
K_1(2\p|q\tm|\rho_2) e^{-2\p i \tm q \t_1} 
\ee
up to the tree-level
$(\rho_2)^2$ term which we discuss below. In fact, 
the $R^4$ terms are accompanied under supersymmetry
by other eight-derivative
terms involving the NS dilaton, which can be simply obtained
by replacing the Riemann tensor by \cite{gross}:
\be
\bar{R}_{\m\n}^{\;\;\;\r\s}=R_{\m\n}^{\;\;\;\r\s}
~-~ {1\over 4} \delta\,_{[\m}^{\,\,[\r} \nabla\raisebox{-0.3ex}{$_{\nu]}$}
\nabla\raisebox{0.3ex}{$^{\sigma]}$} \varphi_{10} 
+\dots
\ee
where we retained terms linear in the dilaton only.
Upon compactification of type IIB string theory to six dimensions on $K3$,
the $\bar{R}^4$ coupling yields 
$\int_{K3} \bar{R}^4 \propto
\chi\,   ( \na\na \ph_{10} )^2$,
where $\chi=24$ is the Euler number of $K3$.
This validates the conjecture in Ref. \cite{dix1} for the
contribution of the D-instantons to the ten-dimensional 
$R^4$ couplings. 

On the other hand, the tree-level term ${24 \zeta(3)\over\p} (\rho_2)^2$
in the $R^4$ coupling (\ref{gg}) does {\it not\/} induce any 
four-derivative dilaton interaction in four dimensions,
in agreement with the absence of the corresponding tree-level
amplitude. This can also be checked by considering the
$\langle T T \tbar\tbar\rangle$ heterotic amplitude
in the large $T_{\rm H}=S_{\rm II}$ limit, which we 
discuss in the Appendix. However, $\Ftilde_1$ 
is defined by Eqs.\ (\ref{ampl3}) and (\ref{ft1h}) up 
to a $T$-dependent integration constant which, as we show in the Appendix,
reproduces in the large $V$ limit the ten-dimensional $(\rho_2)^2$
tree-level term. This  restores the
$Sl(2;\Zint)_{\rho}$ invariance of the $R^4$ couplings in $D=10$.

\section{Concluding Remarks} 
In this work, we analyzed $e^{-1/\lambda}$ non-perturbative
corrections in the context of $N=4$ four-derivative couplings.
These effects should also occur in higher $\Ftilde_g$'s
in the case of $N=2$ compactifications.
A simple counting based on hypermultiplet--vector multiplet 
decoupling shows that these
functions are purely $g$-loop on the heterotic side.
Furthermore, in contrast to the vector multiplet metric
protected by the holomorphicity of the prepotential,
the hypermultiplet metric,
which corresponds to the $\Ftilde_{g=0}$ coupling, 
can receive such corrections in type II (or type I) vacua
\cite{bbs,ov}.
Finally, at the $N=1$ level, we also expect such corrections
to appear in the K\"ahler potential, even in the heterotic theory.
\\[5mm]
\noindent{\bf Acknowledgements} 
 
We are grateful to E. Kiritsis and N. Obers for helpful discussions.
\begin{flushleft}
{\large\bf Appendix}\end{flushleft}
\renewcommand{\theequation}{A.\arabic{equation}}
\renewcommand{\thesection}{A.}
\setcounter{equation}{0}

Here we discuss the amplitude (\ref{aterm}) with $\f_1=\f_2=T$.
Following the same steps as in Section 3, and neglecting the windings
of the two-torus in the large $\iT$ limit, it is easy to show that
Eq.\ (\ref{ampl3}) still holds, so that the
$\langle T T \tbar\tbar\rangle$ amplitude can still be derived
from $\Ftilde_1$ given in Eq. (\ref{ft1h}):
\be
\label{TT}
{\cal A}_{TT}\approx{1\over (T_2)^2}
\left( \pa_T - {i\over 2\iT}\right)\left( \pa_T + {i\over 2\iT}\right) 
\Ftilde_1\ = {1\over (\iT)^2} \pa_T^2 \Ftilde_1 .
\ee
The large $\iT$ expansion of $\Ftilde_1$ 
given in Eqs.\ (\ref{ampl4}), (\ref{h1ii}) and (\ref{amplK})
still holds, up to one additional term coming from 
the contribution of the $\tm^I=0$ terms in the Poisson
resummed formula (\ref{ft11h}):
\be
\iT 
\int_{\F} {d^2 \tau \over(\t_2)^2}  \sum_{q^i}
e^{-\p\t_2 q^t\Mh q ~+~\p i\t_1 q^t\Lh q }
~\etab\ .
\ee 
This term, linear in $\iT$, would be mapped to a 
tree-level contribution proportional to $\iS=e^{-2\ph_4}$
in type IIA and IIB theories.
It is however 
annihilated by the derivative in Eq.\ (\ref{TT}), and therefore
does not contribute to the physical
$\langle T T \tbar\tbar\rangle$ amplitude. It merely
corresponds to a choice of the integration constant in
Eq.\ (\ref{ampl3}).

In order to study the large $K3$ volume limit of the above term,
we again make use of the decomposition (\ref{deck3}) of the
$K3$ moduli matrix $\Mh$. Terms 
with a non-zero 4-cycle charge $q^4$ are exponentially
suppressed (as $O(e^{-V})$) and can be neglected.
After Poisson resummation in the zero-cycle charge $q^0 \to \tilde{q}_0$,
we can distinguish three types of contributions. States
with  $\tilde{q}_0\neq 0$ and 22 two-cycle charges $q'\neq 0$ 
are still suppressed as $O(e^{-\sqrt{V}})$. States with
$\tilde{q}_0=0$ yield a divergent contribution
of order $S_2 \sqrt{V}=e^{-2\ph_{10}} V^{3/2} \iT$ (in type II variables):
\be
\iS \sqrt{V}
\int_{\F} {d^2 \tau \over(\t_2)^{5/2}}  \sum_{q'_i}
e^{-\p\t_2 q'^t\Mh q' ~+~\p i\t_1 q'^t \Lh q' }
~\etab\ ,
\ee
which depends only on the moduli of the unit volume Ricci-flat metric of
$K3$, up to the overall factor $\iS \sqrt{V}$.
Finally, states with vanishing two-cycle charges $q'$ and
non-zero $\tilde{q}_0$ have $\Mh_{00}=1/V$ and yield
\ba
24 \iS \sqrt{V}
 \int_{0}^{\infty} {d\t_2 \over (\t_2)^{5/2}} 
\sum_{\tilde{q}_0\ne 0} e^{-\p\t_2 (\tilde{q}_0)^2/V}
&=& 24 { \iS \Gamma(3/2)\over  \p^{3/2} } \sum_{\tilde{q}_0\ne 0} 
{1\over |\tilde{q}_0|^3}
\cr
&=& {24 \zeta(3) \iS \over \p V}\ .
\ea
Factorizing the $\T^2$ volume and going to the ten-dimensional variable
$(\rho_2)^2=e^{-2\ph_{10}}$ $=S_2/(V \iT)$, we recover the 
type IIB tree-level $(\rho_2)^2$ term of Eq.\ (\ref{gg}). 
The above discussion also applies to the type IIA theory and,
at the same time, we recover the tree-level 
${2 \zeta(3) \over \p} e^{-2\ph_{10}} R^4$ term present in the 
$D=10$ type IIA theory.

\end{document}